\documentclass{article}

\usepackage{theorem}
\usepackage{subfigure}
\usepackage{epsfig}
\usepackage{amssymb}
\usepackage{fullpage}
\usepackage{amsmath}
\usepackage{ifthen}
\usepackage{verbatim}

\newtheorem{theorem}	 			{Theorem}[section]

{\theorembodyfont{\rmfamily} }
{\theorembodyfont{\rmfamily} }
{\theorembodyfont{\rmfamily} }
{\theorembodyfont{\rmfamily} }
{\theorembodyfont{\rmfamily} }
{\theorembodyfont{\rmfamily} \newtheorem{q}			[theorem]
{Question}}
{\theorembodyfont{\rmfamily} }
{\theorembodyfont{\rmfamily} }
\theoremstyle{break}
{\theorembodyfont{\rmfamily} }

\newcommand{\argmax}{\operatornamewithlimits{argmax}}

\newcommand{\prob}[2][]{\text{\bf Pr}\ifthenelse{\not\equal{}{#1}}{_{#1}}{}\!\left[#2\right]}
\newcommand{\expect}[2][]{\text{\bf E}\ifthenelse{\not\equal{}{#1}}{_{#1}}{}\!\left[#2\right]}

\newcommand{\dist}{F}

\newcommand{\disti}[1][i]{{\dist_{#1}}}

\newcommand{\val}{v}
\newcommand{\vals}{{\mathbf \val}}

\newcommand{\vali}[1][i]{{\val_{#1}}}

\newcommand{\vv}{\varphi}
\newcommand{\vvi}[1][i]{{\vv_{#1}}}

\newcommand{\dens}{f}

\newcommand{\densi}[1][i]{{\dens_{#1}}}

\newcommand{\bid}{b}

\newcommand{\bidi}[1][i]{{\bid_{#1}}}

\def\sse{\subseteq}

\def\eps{\epsilon}

\title{Approximately Optimal Mechanism Design:\\ Motivation,
Examples, and Lessons Learned\thanks{This paper is based on
a talk given by the author at the 15th ACM Conference on Economics and
Computation (EC), June 2014.}}
\author{Tim
Roughgarden\thanks{Department of Computer Science,
Stanford University, 462 Gates Building, 353 Serra Mall, Stanford, CA 94305.
Email: {\tt tim@cs.stanford.edu}.}}
\date{\today}

\begin{document}

\maketitle

\section{Introduction}

\subsection{Preamble}

Optimal mechanism design enjoys a beautiful
and well-developed theory, and also a number of killer applications.
Rules of thumb produced by the field influence everything
from how governments sell wireless spectrum licenses to how the major
search engines auction off online advertising.

There are, however, some basic problems for which the
traditional optimal mechanism design approach is ill-suited --- either
because it makes overly strong assumptions, or because it advocates
overly complex designs.  The thesis of this paper is that 
{\em approximately optimal} mechanisms allow us to reason about
fundamental questions that seem out of reach of the traditional theory.

\subsection{Organization}

This survey has three main parts.  The first part reviews a
couple of the greatest hits of optimal mechanism design, the single-item
auctions of Vickrey and Myerson.  
We'll see how
taking baby steps beyond these canonical settings already
highlights limitations of the traditional optimal mechanism design
paradigm, and motivates a more relaxed approach.  
This part also describes the
approximately optimal mechanism design paradigm --- how it works, and
what we aim to learn by applying it.

The second and third parts of the survey cover two case
studies, where we instantiate the general design paradigm to
investigate two basic questions.  In the first example,
we consider revenue maximization in a single-item auction with
heterogeneous bidders.  Our goal is to understand if
complexity --- in the sense of detailed distributional knowledge ---
is an essential feature of good auctions for this problem, or
alternatively if there are simpler auctions that are near-optimal.
The second example considers welfare maximization with multiple items.
Our goal here is similar in spirit: when is complexity --- in the form
of high-dimensional bid spaces --- an essential feature of every auction
that guarantees reasonable welfare?  Are there interesting
cases where low-dimensional bid spaces suffice?

\section{The Optimal and Approximately Optimal Mechanism Design
  Paradigms: Vickrey, Myerson, and Beyond}

\subsection{Example: The Vickrey Auction}

Let's briefly recall the Vickrey or second-price single-item
auction~\cite{V61}. 
Consider a single seller with a single item; assume for simplicity
that the seller has no value for the item.
There are $n$ bidders, and each bidder~$i$ has a valuation $\vali$
that is unknown to the seller.
Vickrey's auction is designed to maximize the welfare, which in a
single-item auction just means awarding the item to the bidder with
the highest valuation.
This sealed-bid auction collects a bid from each bidder, awards the
item to the highest bidder, and charges the second-highest price.
The point of the pricing rule is to ensure that truthful bidding is a
dominant strategy for every bidder.
Provided every bidder follows its dominant strategy, the auction
maximizes welfare ex post (that is, for every valuation profile).

In addition to being theoretically optimal, the Vickrey auction has a
simple and appealing format.  
Plenty of
real-world examples resemble the Vickrey auction.  In light of
this confluence of theory and practice, what else could we ask for?
To foreshadow what lies ahead, we mention that when selling
multiple non-identical items, the generalization of the Vickrey
auction is much more complex.

\subsection{Example: Myerson's Auction}\label{ss:iid}

What if we want to maximize the seller's revenue rather than the
social welfare?  Since there is no single auction that maximizes
revenue ex post, the standard approach here is to maximize the
expected revenue with respect to a prior distribution over bidders'
valuations.  So, assume bidder $i$'s valuation is drawn independently from a
distribution $\disti$ that is known to the seller.  For the moment,
assume also that bidders are homogeneous, meaning that their
valuations are drawn i.i.d.\ from a known distribution $\dist$.

Myerson~\cite{M81} identified the optimal auction in this context, which
is a simple twist on the Vickrey auction --- a
second-price auction with a reserve price~$r$.\footnote{That is, the
  winner is the highest bidder with bid at least $r$, if any.  If
  there is a winner, it pays either the reserve price or the
  second-highest bid, whichever is larger.}  Moreover, the optimal
reserve price is simple and intuitive --- it is just the {\em monopoly
  price} $\argmax_p [p \cdot (1-\dist(p))]$ for the distribution $F$,
the optimal take-it-or-leave-it offer to a single
bidder with valuation drawn from $\dist$.  Thus, to implement the
optimal auction, you don't need to know much about the valuation
distribution $\dist$ --- just a single statistic, its monopoly price.

Once again, in addition to being theoretically optimal, Myerson's
auction is simple and appealing.  It is more or less equivalent to an
eBay auction, where the reserve price is implemented using an opening bid.
Given this success,
why do we need to enrich the traditional optimal mechanism design paradigm?
As we'll see, when bidders' valuations are not i.i.d., the
theoretically optimal auction is much more complex and no longer
resembles the auction formats that are common in practice.

\subsection{The Optimal Mechanism Design Paradigm}

Having reviewed two famous examples, let's zoom out and be more
precise about the optimal mechanism design paradigm.
The first step is to identify the design space of possible mechanisms,
such as the set of all sealed-bid auctions.  The second step is to 
specify some desired properties.  In this talk, we focus only on cases
where the goal is to optimize some objective function that has
cardinal meaning, and for which relative approximation makes sense.
We have in mind objectives such as the seller's revenue (in expectation
with respect to a prior) or social welfare (ex post) in a transferable
utility setting.  The goal of the analyst is then to identify one or
all points in the design space that possess the desired properties ---
for example, to characterize the mechanism that maximizes the welfare
or expected revenue.

What can we hope to learn by applying this framework?  The traditional
answer is that by solving for the optimal mechanism, we hope to
receive some guidance about how to solve the problem.  With the
Vickrey and Myerson auctions, we can take the theory quite literally
and simply implement the mechanism advocated by the theory.
More broadly, one looks for features present in the theoretically
optimal mechanism that seem broadly useful --- for example, Myerson's
auction suggests that combining welfare maximization with suitable
reserve prices is a potent approach to revenue-maximization.

There is a second, non-traditional answer that we exploit
explicitly when we extend the paradigm to accommodate approximation.
Even when the theoretically optimal mechanism is not directly useful
to the practitioner, for example because it is too complex, it is
directly useful to the analyst.  The reason is that the performance
of the optimal mechanism can serve as a benchmark, a yardstick against
which we measure the performance of other designs that are more
plausible to implement.

\subsection{The Approximately Optimal Mechanism Design Paradigm}\label{ss:approxwhy}

To study approximately optimal mechanisms, we again begin with a
design space and an objective function.  Often the design space is
limited by side constraints such as a ``simplicity'' constraint.  For
example, we later consider mechanisms with limited distributional
knowledge, and those with low-dimensional bid spaces.

The new ingredient of the paradigm is a {\em benchmark}.  This is a
target objective function value that we would be ecstatic to
achieve.  Generally, the working hypothesis will be that no mechanism
in the design space realizes the full value of the benchmark, so the
goal is to get as close to it as possible.
In the two examples we discuss, where the design space is limited by a
simplicity constraint, a simple and natural benchmark is the performance
achieved by an unconstrained, arbitrarily complex mechanism.
The goal of the analyst is to identify a mechanism in the design space
that approximates the benchmark as closely as possible.
For example, it is clearly interesting to establish that there is a
``simple'' mechanism with performance almost as good as an arbitrarily
complex one.

What is the point of applying this design paradigm?  The first goal is
exactly the same as with the traditional optimal mechanism design
paradigm.  Whenever you have a principled way of selecting out one
mechanism from many, you can hope that the distinguished mechanism is
literally useful or highlights features that are essential to
good designs.  The approximation paradigm provides a novel way to
identify candidate mechanisms.

There is a second reason to use the approximately optimal mechanism
design paradigm, which has no analog in the traditional approach.
The approximation framework enables the analyst to quantify the cost
of imposing side constraints on a mechanism design space.  For
example, if there is a simple mechanism with performance close to
that of the best arbitrarily complex mechanism, then this fact suggests
that simple solutions might be good enough.  Conversely, if every
point in the design space is far from the benchmark, then this
provides a forceful argument that complexity is an essential feature
of every reasonable solution to the problem.

\subsection{Two Case Studies}

Sections~\ref{s:rev} and~\ref{s:wel} instantiate the approximately
optimal mechanism design paradigm to study two fundamental questions.
We first study expected revenue-maximization in single-item auctions,
with bidders that have independent but not necessarily identically
distributed valuations.
The theoretically optimal mechanism can be complex, in the sense
that it requires detailed distributional knowledge.
We use the approximation paradigm to identify when such complexity
is an inevitable property of every near-optimal auction.

Our second case study concerns welfare maximization.  Here, the
complexity stems from selling multiple non-identical items.
Again, the theoretically optimal mechanism is well known but suffers
from several drawbacks that preclude direct use.  We apply the
approximation paradigm to identify when simpler mechanisms, meaning
mechanisms with low-dimensional bid spaces, can perform well, versus when
complex bid spaces are necessary for non-trivial welfare guarantees.

\subsection{Many Applications of the Approximation Paradigm}\label{ss:refs}

An enormous amount of research over the past fifteen years, largely
but not entirely in
the computer science literature, can be viewed as instantiations of
the approximately optimal mechanism design paradigm.  
This paper merely singles out two recent examples that are near and
dear to the author's heart.

For example, all
of the following questions have been studied through the lens of
approximately optimal mechanisms.
\begin{enumerate}

\item What is the cost of imposing bounded communication in
  settings with very large type spaces, such as combinatorial
  auctions?  This line of research originated 
  in~\cite{NS06} and is surveyed in~\cite{comm_survey}.

\item What is the cost of imposing bounded computation in settings
  that involve computationally difficult optimization problems, such
  as combinatorial auctions?  Two early papers are~\cite{LOS02,NR99}
  and a recent survey is~\cite{amd_survey}.

\item What is the cost of limiting the distributional knowledge of a
  mechanism?  
Several papers in the economics
  literature~\cite{BV03,BK96,N03,S03} shed light on this question.
The approximation interpretation is explicit in~\cite{DRY10}; see
also~\cite[Chapter 4]{hartline} for a survey.  
The case study in Section~\ref{s:rev} is another example of work in
this vein.

\item Are there auctions that achieve good revenue in the worst case
  (i.e., ex post)?  This question was formalized using the
  approximately optimal mechanism design framework in~\cite{G+06};
  see~\cite[Chapter 5]{hartline} for a recent survey.

\item Are there mechanisms with ``simple'' allocation rules that perform
  almost as well as arbitrarily complex mechanisms?  
For example, see~\cite{CHK07,HR09} for revenue guarantees for auctions
that make use only of welfare-maximization supplemented by reserve
prices.

\item Are there mechanisms with ``simple'' pricing rules that perform
  almost as well as arbitrarily complex mechanisms?  
See~\cite{LB10,C+14} for case studies in combinatorial and keyword
auctions, respectively.
The case study in Section~\ref{s:wel} is another example of
this and the preceding directions.

\end{enumerate}

\section{Case Study: Do Good Single-Item Auctions Require Detailed
  Distributional Knowledge?}\label{s:rev}

This section applies the approximately optimal mechanism design
paradigm to the problem of revenue-maximization in single-item auctions.
The take-away from this exercise is that the amount of
distributional knowledge required for near-optimal revenue is governed 
by the degree of bidder heterogeneity.

\subsection{Optimal Single-Item Auctions}

We now return to expected revenue-maximization in single-item
auctions, but allow {\em heterogeneous} bidders, meaning that
each bidder $i$'s private valuation $\vali$ is drawn independently
from a distribution $\disti$ that is known to the seller.
Myerson~\cite{M81} characterized the optimal auction, as a function of
the distributions $\dist_1,\ldots,\dist_n$.

The trickiest step of Myerson's optimal auction is the first one,
where each bid $\bidi$ is transformed into a {\em virtual bid}
$\vvi(\bidi)$, defined by
\begin{equation}\label{eq:vv}
\vvi(\bidi) = \bidi - \frac{1-\disti(\bidi)}{\densi(\bidi)}.
\end{equation}
The exact functional form in~\eqref{eq:vv} is not important for this
paper, except to notice that computing $\phi_i(\bidi)$ requires
knowledge of the distribution, namely of $\densi(\bidi)$ and $\disti(\bidi)$.

Given this transformation, the rest of the auction is
straightforward.  The winner is the bidder with the highest positive
virtual bid (if any).  To make truthful bidding a dominant strategy,
the winner is charged the minimum bid at which it would continue to be
the winner.\footnote{We have only described the optimal auction in the
  special case where each distribution $\disti$ is {\em regular},
  meaning that the virtual valuation functions $\vvi$ are
  nondecreasing.  The general case ``monotonizes'' the virtual
  valuation functions --- monotonicity is essential for
  incentive-compatibility --- and then applies the same three
  steps~\cite{M81}.}

When all the distributions $\disti$ are equal to a common~$\dist$, 
and hence all virtual valuation functions $\vvi$ are identical, the
optimal auction simplifies and is simply a second-price auction with a
reserve price of $\vv^{-1}(0)$, which turns out to be the monopoly
price for $\dist$.
In this special case, the optimal auction requires only modest
distributional knowledge --- a singe statistic, the monopoly price.
In general, the optimal auction does not simplify further than the
description above, and detailed distributional knowledge is required
to compute and compare the virtual bids of bidders with different
valuation distributions.

\subsection{Motivating Question}

This section uses the approximately optimal mechanism design paradigm
to study the following question.
\begin{itemize}

\item [] {\em Does a near-optimal single-item auction require
  detailed distributional knowledge?}

\end{itemize}

To study this question formally, we need to parameterize the ``amount
of knowledge'' that the seller has about the valuation distributions.
We look to computational learning theory, a
well-developed branch of computer science~\cite{V84}, for inspiration.
We consider a seller that does not know the valuation distributions
$F_1,\ldots,F_n$, except inasmuch as it knows $s$ valuation profiles
$\vals^{(1)},\ldots,\vals^{(s)}$ that have been sampled i.i.d.\ from
these distributions.  In an auction context, an obvious
interpretation of these samples is as the valuations of comparable
bidders in past auctions for comparable items, as inferred from bid
data.  See Ostrovsky and Schwarz~\cite{OS09} for a real-world
example of this approach, in the context of setting reserve prices in
Yahoo! keyword auctions.

Thus, our design space is the set of auctions that depend on the
valuation distributions only through samples.  Formally, for a
parameter $s \ge 1$, a point in the design space is a function from $s$
valuation profiles (the samples) to a single-item auction, which is
then run tomorrow on a fresh valuation profile drawn from $\dist_1 \times
\cdots \times \dist_n$.  Our objective function is the expected
revenue, where the expectation is over both the samples (which
determines the auction used) and the final valuation profile (which
determines the revenue earned by the chosen auction).

Our benchmark --- the highest expected revenue we could conceivably
obtain --- is simply the expected revenue earned by Myerson's optimal
auction for the distributions $\dist_1,\ldots,\dist_n$.  
We call this the {\em Myerson benchmark}.
Thus, we are comparing the optimal expected revenue obtainable by a
seller with partial distributional knowledge to that by a seller with
full distributional knowledge.
The goal is to understand the amount of knowledge (i.e., the number of
samples) needed to earn expected revenue at least
$(1-\eps)$ times the Myerson benchmark, where $\eps$ is a parameter such as
$0.1$ or $0.01$.

\subsection{Formalism: One Bidder}

\begin{figure}
\begin{center}
\includegraphics[scale=.7]{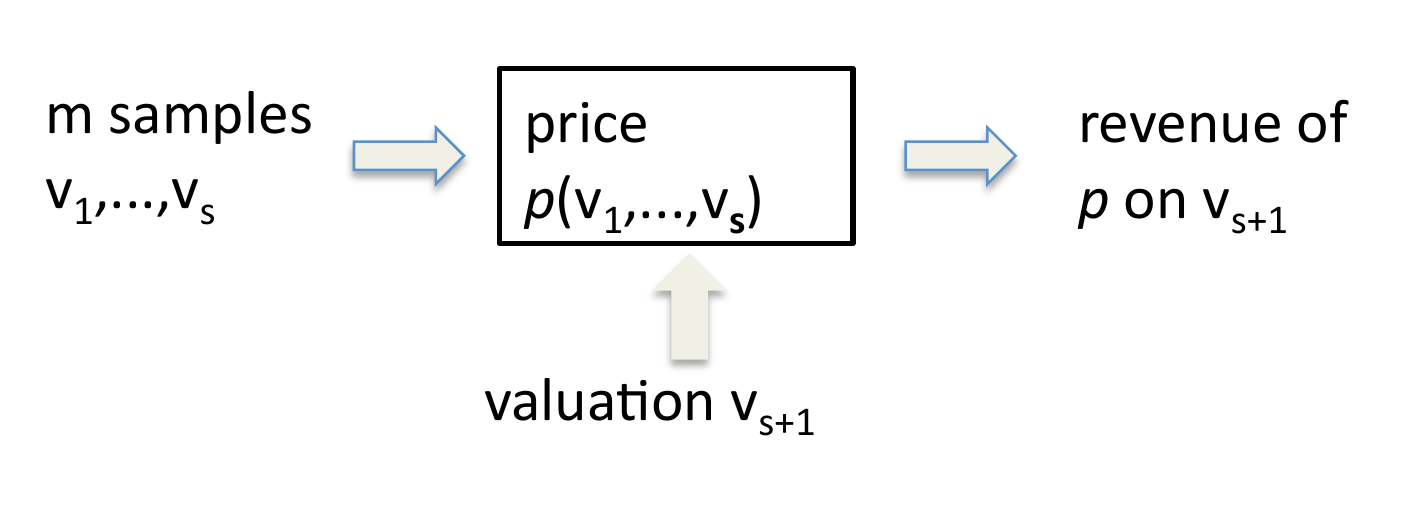}
\caption{Single-bidder formalism.  A take-it-or-leave-it offer $p$ is
  chosen as a function of $s$ i.i.d.\ samples $\val_1,\ldots,\val_s$ from an
  unknown distribution $\dist$, and applied to a fresh sample
  $\val_{s+1}$ from the same distribution.  The goal is to understand,
  given a parameter $\eps > 0$,
  how many samples $s$ are needed so that there exists a price
  function with expected revenue at least $1-\eps$ times that of the
monopoly price for $\dist$.}
\label{f:single}
\end{center}
\end{figure}

To make sure that the formalism is clear, let's warm up with a simple
example.  In addition to only one seller with one item, suppose
there is also only one bidder, with valuation drawn from a
distribution $\dist$ unknown to the seller.  With only one bidder,
auctions are merely take-it-or-leave-it offers.\footnote{Probability
  distributions over take-it-or-leave-it-offers are also allowed.
  We discuss only deterministic auctions for simplicity of
  presentation, but the results of this section also apply to randomized
  auctions.}
The goal is to design
a function $p(\val_1,\ldots,\val_s)$ from samples
$\val_1,\ldots,\val_s \sim \dist$ to prices that, for every $\dist$,
achieves expected revenue 
$\expect[\val_1,\ldots,\val_s]{p(\val_1,\ldots,\val_s) \cdot
  (1-\dist(p(\val_1,\ldots,\val_s)))}$
close 
to that achieved by the monopoly price $\argmax_p p \cdot
(1-\dist(p))$ of $\dist$.  
In other words, given data from $s$ past transactions, the goal is to
set a near-optimal price for a new bidder encountered tomorrow.
See also Figure~\ref{f:single}.

\subsection{Results for a Single Bidder}\label{ss:singlesample}

We next state a series of results for the single-bidder special case. 
These results are not the main point of this case study, and instead
serve to calibrate our expectations for what might be possible
for single-item auctions with multiple bidders.

The bad news is that, without any assumptions about the unknown
distribution $\dist$, 
no finite number of samples yields a non-trivial expected revenue
guarantee for every $\dist$.  That is, for every finite $s$, there is
a valuation distribution $\dist$ such that you learn essentially
nothing about $\dist$ from $s$ samples.\footnote{For example, for a
  parameter $M \rightarrow \infty$, consider distributions $\dist$
  that put a point mass of $1/M$ at $M$ and are otherwise zero.}
This observation motivates restrictions on the unknown distribution.

The good news is that under a standard ``regularity'' condition, intuitively
stating that the tail of $\dist$ is no heavier than a power-law
distribution, is sufficient for interesting positive
results.\footnote{Formally, a distribution~$\dist$ is {\em regular} if
  its virtual valuation distribution~\eqref{eq:vv} is nondecreasing.}
Even just {\em one} sample can be used to obtain a non-trivial revenue
guarantee for unknown regular distributions: 
for every such $\dist$, the function $p(\val_1) =
\val_1$ --- using yesterday's bid as tomorrow's price ---
yields expected revenue at least 50\% times that of the
monopoly price.\footnote{This is a consequence of the following
  special case of the Bulow-Klemperer 
  theorem on auctions vs.\ negotiations~\cite{BK96}: the expected
  revenue of a Vickrey 
  auction with two bidders with valuations drawn i.i.d.\ from a
  regular distribution $\dist$ is at least that of an optimal auction
  (i.e., the monopoly price) for a single bidder with valuation drawn
  from $\dist$.}

What if we want a better revenue guarantee, like 90\% or 99\% of 
this benchmark?  To achieve a $(1-\eps)$-approximation
guarantee, we expect 
the number of samples required to increase with $1/\eps$.  
Happily, the amount of data required is relatively modest, scaling as a
polynomial function of $1/\eps$.  For an unknown regular distribution,
this function is roughly~$\eps^{-3}$~\cite{DRY10,HMR14}.
The sample complexity improves if we impose stronger conditions on the
tails of the valuation distribution.  For example, if $\dist$
satisfies the monotone hazard rate condition --- meaning
$\dens(x)/(1-\dist(x))$ is nondecreasing with $x$ ---- then roughly
$\eps^{-3/2}$ samples are necessary and sufficient to achieve a
$(1-\eps)$-approximation of the benchmark~\cite{HMR14}.
The upper bounds on sample complexity follow from natural pricing
strategies, such as choosing the monopoly price for the empirical
distribution of the samples.

\subsection{Formalism: Multiple Bidders}

\begin{figure}
\begin{center}
\includegraphics[scale=.7]{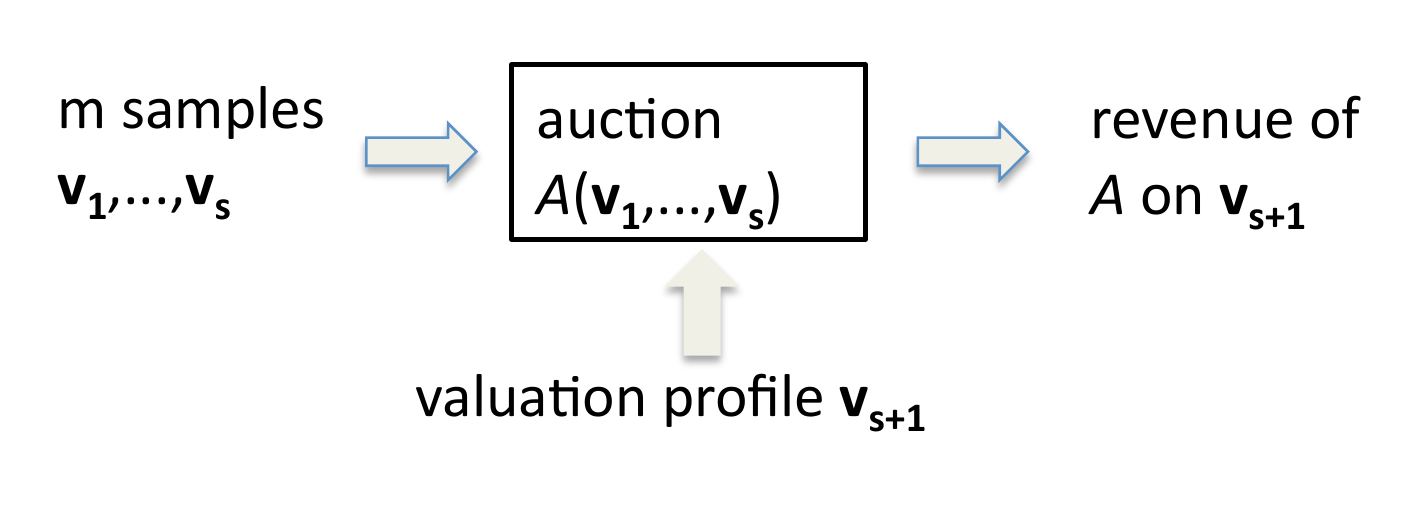}
\caption{Multiple-bidder formalism.
A single-item auction $A$ is
  chosen as a function of $s$ i.i.d.\ samples $\vals_1,\ldots,\vals_s$ from an
  unknown distribution $\dist_1 \times \cdots \times \dist_n$, and
  applied to a fresh sample 
  $\vals_{s+1}$ from the same distribution.  
The benchmark is the expected revenue of the Myerson-optimal auction
for $\dist_1,\ldots,\dist_n$.}
\label{f:multiple}
\end{center}
\end{figure}

Generalizing the formalism to single-item auctions with multiple
bidders proceeds as one would expect.
The seller is now given $s$ samples $\vals_1,\ldots,\vals_s$, where
each sample $\vals_j$ is a valuation profile, comprising one valuation
(drawn from $\disti$) for each bidder $i$.
The seller picks an auction $A(\vals_1,\ldots,\vals_s)$ that is a
function of these samples only.
Recall that the Myerson benchmark is
the expected revenue of the
optimal auction for $\dist_1,\ldots,\dist_n$. 
The goal is to design
a function $A(\val_1,\ldots,\val_s)$ from samples
$\vals_1,\ldots,\vals_s \sim \dist_1 \times \cdots \times \dist_n$ to
single-item auctions that, for every $\dist_1,\ldots,\dist_n$,
achieves expected revenue close to this benchmark.
As in the single-bidder case, the expectation is over both the past bid
data (the samples) and the bidders (a fresh sample from the same
distributions).
See also Figure~\ref{f:multiple}.

\subsection{Positive Results}

The hope is that our positive results for the single-bidder problem
(Section~\ref{ss:singlesample}) 
carry over to single-item auctions with multiple bidders. 
First, provided $\dist_1,\ldots,\dist_n$ are regular distributions,
it is still possible to get a coarse but non-trivial
approximation (namely, 25\%) with a single sample --- this follows
from a generalization of the Bulow-Klemperer theorem given
in~\cite{HR09}.  But what about very close approximations, like 90\%
or 99\%?

In the special case where bidders are homogeneous --- meaning have
identically distributed valuations --- the positive results for a
single bidder continue to hold.  Intuitively, the reason is that the
form of the optimal auction is independent of the number of bidders
--- it is simply a second-price auction with a reserve set to the
monopoly price for the distribution $\dist$.  Since a single statistic
about the distribution $\dist$ determines
the optimal auction for an arbitrary number of homogeneous bidders, it
makes sense that the sample complexity of approximating this optimal
auction is independent of $n$. 

Thus, in these cases, the amount of data --- the granularity of
knowledge about the valuation distributions --- necessary for
near-optimal revenue is relatively modest,
and does not depend on the number of bidders.

\subsection{Negative Results}\label{ss:lb}

The approximately optimal mechanism design paradigm identifies a
qualitative difference between the cases of homogeneous and
heterogeneous bidders.  When bidders are heterogeneous and we seek a
close approximation of the optimal revenue, the sample complexity
depends fundamentally on the number of bidders.
\begin{theorem}[\cite{CR14}]\label{t:cr14}
There is a constant $c > 0$ such that, for every sufficiently small
$\eps > 0$ and every $n \ge 2$, there is no auction that depends on
at most $cn/\sqrt{\eps}$ samples and has expected revenue at least $1-\eps$
times the Myerson benchmark for every profile
$\dist_1,\ldots,\dist_n$ of regular distributions.
\end{theorem}
The valuation distributions used in the proof of Theorem~\ref{t:cr14}
are not pathological --- exponential distributions, truncated
at different maximum values, already yield the lower bound.

The proof of Theorem~\ref{t:cr14} shows more generally that every
auction that fails to implicitly learn all bidders' virtual valuation
functions (recall~\eqref{eq:vv}) up to small error is doomed to having
expected revenue less than $(1-\eps)$ times the Myerson benchmark in
some cases.  In this sense, detailed knowledge of the valuation
distributions is an unavoidable feature of every near-optimal
single-item auction with heterogeneous
bidders.\footnote{There is also a converse to Theorem~\ref{t:cr14}:
  for every $\eps > 0$ and $n \ge 1$, and for an arbitrary number of
  bidders with $n$ distinct valuation distributions, a polynomial
  number (in
  $n$ and $\eps^{-1}$) of samples is sufficient to achieve a
  $(1-\eps)$-approximation of the Myerson benchmark~\cite{CR14}.}

\section{Case Study: Do Good Combinatorial Auctions Require Complex
  Bid Spaces?}\label{s:wel}

In this section we switch gears and study 
the problem of allocating multiple items to bidders with private
valuations to maximize the social welfare.
We instantiate the approximately optimal mechanism design
paradigm to identify conditions on bidders' valuations that are
necessary and sufficient for the existence of simple combinatorial
auctions.
The take-away from this section is that rich bidding spaces are
an essential feature of every good combinatorial auction when items
are complements, while simple auctions can perform well when bidders'
valuations are complement-free.

\subsection{The VCG Mechanism}

We adopt the standard setup for allocating multiple items via a
combinatorial auction.  There are $n$ bidders and $m$ non-identical
items.  Each bidder has, in principle, a different private valuation
$\vali(S)$ for each bundle~$S$ of items it might receive.  Thus,
each bidder has $2^m$ private parameters.  In this section, we assume
that the objective is to determine an allocation $S_1,\ldots,S_n$ that
maximizes the social welfare $\sum_{i=1}^n \vali(S_i)$.

The Vickrey auction can be extended to the case of multiple items;
this extension is the Vickrey-Clarke-Groves (VCG)
mechanism~\cite{V61,C71,G73}.  The VCG mechanism is a
direct-revelation mechanism, so each bidder $i$ reports a valuation
$\bidi(S)$ for each bundle of items~$S$.  The mechanism then computes an
allocation that maximizes welfare with respect to the reported
valuations.  As in the Vickrey auction, suitable payments make truthful
revelation a dominant strategy for every bidder.

Even with a small number of items, the VCG mechanism is a non-starter
in practice, for a number of reasons~\cite{lovely}.  We focus here on the
first step.  Every direct revelation mechanism, including the VCG
mechanism, solicits $2^m$ numbers from each bidder.  This is an
exorbitant number: roughly a
thousand parameters when $m=10$, roughly a million when $m=20$.

\subsection{Motivating Question}

In this case study, we apply the approximately optimal mechanism
design paradigm to study the following question.
\begin{itemize}

\item [] {\em Does a near-optimal combinatorial auction require rich
  bidding spaces?}

\end{itemize}
Thus, as in the previous case study, we seek conditions
under which ``simple auctions'' can ``perform well.''
This time, our design space of ``simple auctions'' consists of mechanism
formats in which 
the dimension of every player's bid space is growing polynomially with
the number $m$ of items (say $m$ or $m^2$), rather than exponentially
with $m$ as in the VCG mechanism.

``Performing well'' means, as usual, achieving objective function value
(here, social welfare) close to that of a benchmark.
We use the {\em VCG benchmark}, meaning the welfare obtained by the
best arbitrarily complex mechanism (the VCG mechanism), which is
simply the maximum-possible social welfare.

This case study contributes to the debate about whether or not
package bidding 
is an important feature of combinatorial auctions, a topic
over which much blood and ink has been spilled over the past twenty
years.
We can identify auctions with no or limited packing bidding with
low-dimensional mechanisms, and those that support rich package
bidding with high-dimensional mechanisms.
With this interpretation, our results make precise the intuition that
flexible package bidding is crucial when items are complements, but
not otherwise.

\subsection{A Simple Auction: Selling Items Separately}

Our goal is to understand the power and limitations of the entire
design space of low-dimensional mechanisms.  
To make this goal more concrete, we begin by examining a
specific simple auction format.

The simplest way of selling multiple items is by selling each
separately.  Several specific auction formats implement this general
idea.  We analyze one such format, simultaneous first-price
auctions~\cite{B99}.  In this auction, each bidder submits
simultaneously one bid per item --- only $m$ bidding parameters,
compared with its $2^m$ private parameters --- and each item is sold
in parallel using a first-price auction.

When do we expect simultaneous first-price auctions to have reasonable
welfare at equilibrium?
Not always.  With general bidder valuations, and in particular when
items are complements, we might expect severe inefficiency due to
the ``exposure problem'' (e.g.,~\cite{milgrom}).  For example,
consider a bidder in an 
auction for wireless spectrum licenses that has large value for full
coverage of California but no value for partial coverage.
When items are sold separately, such a bidder has no vocabulary to
articulate its preferences, and runs the risk of obtaining a subset of
items for which it has no value, at a significant price.

Even when there are no complementarities amongst the items, we expect
inefficiency when items are sold separately (e.g.,~\cite{krishna}).
The first reason is 
``demand reduction,'' where a bidder pursues fewer items than it truly
wants, in order to obtain them at a cheaper price.
Second, if bidders' valuations are drawn independently from different
valuation distributions, then even with a single item,
Bayes-Nash equilibria are not always fully efficient.

\subsection{Valuation Classes}

\begin{figure}
\begin{center}
\includegraphics[scale=.5]{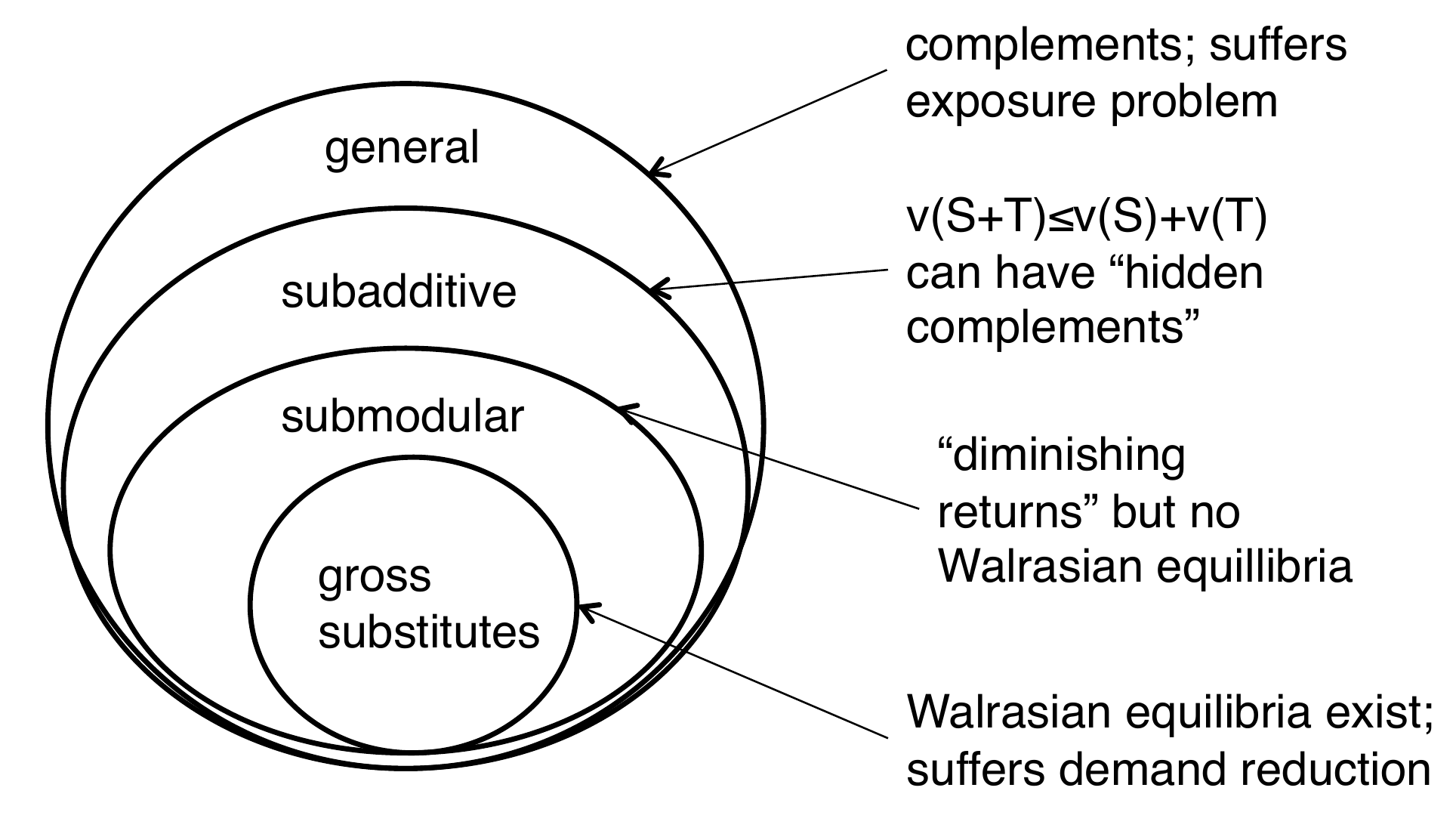}
\caption{Valuation classes.}
\label{f:vals}
\end{center}
\end{figure}

Our discussion so far suggests that simultaneous first-price auctions
are unlikely to work well with general valuations, and suffer from
some degree of inefficiency even with simple bidder valuations.
To parameterize the performance of this auction format, we
introduce a hierarchy of bidder valuations (Figure~\ref{f:vals});
the literature also considers more fine-grained
hierarchies~\cite{feige,LLN01}.

The biggest set corresponds to general valuations, which can encode
complementarities among items.
The other three sets denote different notions of
``complement-free'' valuations.
In this survey, we focus on the most permissive of these, {\em
  subadditive valuations}.  Such a valuation $\vali$ is monotone
($\vali(T) \sse \vali(S)$ whenever $T \sse S$) and satisfies $\vali(S
\cup T) \le \vali(S) + \vali(T)$ for every pair $S,T$ of
bundles.
This class is significantly larger than the well-studied classes of
gross substitutes and submodular valuations.\footnote{Submodularity is 
  the set-theoretic analog of ``diminishing returns:'' $\vali(S \cup
  \{j\}) - \vali(S) \le \vali(T \cup \{j\}) - \vali(T)$ whenever $T
  \sse S$ and $j \notin S$.  The gross substitutes condition ---
  which states that a bidder's demand for an item only increases as the
  prices of other items rise --- is strictly stronger and guarantees
  the existence of Walrasian equilibria~\cite{KC82,GS99}.}
In particular, subadditive   valuations can have ``hidden
complements'' --- meaning two items become complementary given
that a third item has already been acquired --- while submodular
valuations cannot~\cite{LLN01}.

\subsection{When Do Simultaneous First-Price Auctions Work Well?}\label{ss:s1a}

Our intuition about the performance of simultaneous first-price
auctions translates nicely into rigorous statements.  First, for
general valuations, selling items separately can be a disaster.
\begin{theorem}[\cite{H+11}]\label{t:h+11}
With general bidder valuations, simultaneous first-price auctions can
have mixed-strategy Nash equilibria with expected welfare arbitrarily
smaller than the VCG benchmark.
\end{theorem}
For example, equilibria of simultaneous first-price auctions need not
obtain even 1\% of the maximum-possible welfare when there are
complementarities between many items.

On the positive side, even for the most permissive notion of
complement-free valuations --- subadditive valuations --- simultaneous
first-price auctions suffer only bounded welfare loss.
\begin{theorem}[\cite{FFGL13}]\label{t:ffgl13}
If every bidder's valuation is drawn independently from a distribution
over subadditive valuations, then the expected welfare obtained at
every Bayes-Nash equilibrium of simultaneous first-price auctions
is at least 50\% of the expected VCG benchmark value.
\end{theorem}
In Theorem~\ref{t:ffgl13}, the valuation distributions of different
bidders do not have to be identical, just independent.
The guarantee improves to roughly 63\% for the special case of
submodular bidder valuations~\cite{ST13}.

Taken together, Theorems~\ref{t:h+11} and~\ref{t:ffgl13} suggest that
simultaneous first-price auctions should work reasonably well if and
only if there are no complementarities among items.

\subsection{Digression on Approximation Ratios}

Before proceeding to our final set of technical results, we pause to
emphasize how worst-case approximation results like
Theorems~\ref{t:h+11} and~\ref{t:ffgl13} should be interpreted.
Many researchers have a tendency to fixate unduly on and take too
literally such approximation guarantees.

Both of the primary
motivations for applying the approximately optimal mechanism design
paradigm strive for qualitative insights, not fine-grained
performance predictions (recall Section~\ref{ss:approxwhy}).
The first goal is to identify mechanisms or mechanism features that are
potentially useful in practice.
The auction formats implicitly recommended by our case studies, such
as selling items seperately with first-price auctions provided
bidders' valuations are sufficiently simple, corroborate well with
folklore beliefs.
The second goal of the approximation paradigm is to quantify the cost
of a side constraint like ``simplicity'' on the mechanism design
space.  
In our case studies, we are coarsely classifying such constraints as
``tolerable'' or ``intolerable'' according to whether or not imposing
the constraint reduces the achievable performance by a modest constant
factor.  This viewpoint leads to interesting and sensible conclusions
in both of our case studies: 
complexity is unavoidable in near-optimal revenue-maximizing
single-item auctions if and only if bidders are 
heterogeneous, and complexity is unavoidable in near-optimal
welfare-maximization auctions for selling multiple items if and only
if there are complementarities among the items.

To the reader who insists on interpreting approximation guarantees
literally, against our advice, we offer a few observations.
First, in most applications of the approximately optimal mechanism
design framework, the benchmark is constructed so that there is
no mechanism in the design space that always achieves 100\% of the
benchmark.  When 100\% is unachievable, the best-possible
approximation is going to be some number bounded below 100\% ---
it cannot be arbitrarily close to 100\% when nothing is tending to
infinity.\footnote{In some cases it makes
  sense to speak of the asymptotic optimality of a mechanism, such
  as the sample complexity results of Section~\ref{s:rev} and in large
  markets (e.g.,~\cite{S03,swinkels}).  Asymptotic results are clearly
  interesting, but are applicable to only a fraction of the problems that
we want to reason about.}
Examples that demonstrate mechanism suboptimality are often ``small''
in some sense, which translates to impossibility results for
worst-case approximation guarantees better than relatively modest
fractions like 50\% or, if you're lucky, 75\%.
Finally, remember that the benchmark being approximated
--- for example, the performance of a mechanism so complex as to be
unrealizable --- is generally not an option on the table.
The benchmark represents a utopia that exists only in the analyst's mind
--- like your favorite baseball team winning 162 games, or receiving
referee reports on your journal submission in less than six months.

Of course, like any general analysis framework, the approximation
paradigm can be abused and should be applied with good taste.  
In settings where the approximately optimal mechanism design paradigm
does not give meaningful results, the approach should be modified ---
by defining a different benchmark, changing the notion of benchmark
approximation, or using a completely different analysis framework.

\subsection{Negative Results}

We now return to the question of when simple mechanisms, meaning
mechanisms with low-dimensional bid spaces, can achieve non-trivial
welfare guarantees.  Section~\ref{ss:s1a} considered the special case of
simultaneous first-price auctions; here we consider the full design
space.

First, the poor performance of simultaneous first-price auctions 
with general bidder valuations is
not an artifact of the specific format: {\em every} simple mechanism
is vulnerable to arbitrarily large welfare loss when there are
complementarities among items.  This impossibility result argues
forcefully for a rich bidding language, such as flexible package
bidding, in such environments.
\begin{theorem}[\cite{R14}]\label{t:r14a}
With general bidder valuations, no family of simple mechanisms
guarantees equilibrium welfare at least a constant fraction of the
VCG benchmark.
\end{theorem}
In Theorem~\ref{t:r14a}, the mechanism family is parameterized by the
number of items $m$; ``simple'' means that the number of dimensions in
each bidder's bid space is bounded above by some polynomial function of
$m$.  The theorem states that for every such family and constant $c >
0$, for all sufficiently large $m$, there is a valuation profile and
a full-information mixed Nash equilibrium of the mechanism with
expected welfare less 
than $c$ times the maximum possible.\footnote{Technically,
  Theorem~\ref{t:r14a} proves this statement for an $\eps$-approximate
  Nash equilibrium --- meaning every player mixes only over strategies
  with expected utility within $\eps$ of a best response --- where
  $\eps > 0$ can be made arbitrarily small.  The same comment applies
  to Theorem~\ref{t:r14b}.}

We already know from Theorem~\ref{t:ffgl13} that, in contrast, simple
auctions can have non-trivial welfare guarantees with complement-free
bidder valuations.  Our final result states that no simple mechanism
outperforms simultaneous first-price auctions with these bidder valuations.
\begin{theorem}[\cite{R14}]\label{t:r14b}
With subadditive bidder valuations, no family of simple mechanisms
guarantees equilibrium welfare more than 50\% of the
VCG benchmark.
\end{theorem}

\section{Conclusions}

\subsection{Motivating Questions Revisited}

To close the circle, we return to the motivating questions
of our case studies and review the answers provided by
the approximately optimal mechanism design paradigm.  The first
question was:
\begin{itemize}

\item [] {\em Does a near-optimal single-item auction require
  detailed distributional knowledge?}

\end{itemize}
To answer this question, we took the design space to be auctions with
limited knowledge of the valuation distributions --- in the form of
$s$ i.i.d.\ samples --- and studied the number of samples
necessary and sufficient to achieve a $(1-\eps)$-approximation of the
Myerson benchmark.
We discovered that the amount of knowledge (i.e., samples) required
scales linearly with the number of distinct valuation distributions
represented in the bidder population.  Thus, detailed distributional
knowledge is required for near-optimal revenue maximization if and
only if the bidders are heterogeneous.

The second motivating question was:
\begin{itemize}

\item [] {\em Does a near-optimal combinatorial auction require rich
  bidding spaces?}

\end{itemize}
Here, we defined the design space to be families of mechanisms for
which the number of parameters in a bidder's bid space grows
polynomially with the number $m$ of items.
We adopted the VCG benchmark, which equals the maximum-possible
social welfare.
We discovered that high-dimensional bid spaces are fundamental to
non-trivial welfare guarantees when there are complementarities among
items, but not otherwise.  We also learned that, in some cases, selling
items separately with first-price auctions achieves the best-possible
worst-case approximation guarantee of any family of simple mechanisms.

\subsection{Further Discussion}

We showed how the approximately optimal mechanism design paradigm
yields basic insights about two fundamental problems.
Moreover, it is not clear how to glean these insights without
resorting to an analysis framework that incorporates approximation.
Our first case study fundamentally involved suboptimality --- the less
knowledge the seller has, the less revenue it can obtain.
Similarly, inefficiency was an unavoidable aspect of our second case
study, since simple mechanisms are suboptimal even in very simple
settings (e.g., due to demand reduction or bidder asymmetry).
An approximation framework is the obvious way to reason about and
compare different degrees of suboptimality.

An alternative idea, given a design space and an objective function,
is to simply identify the mechanism in the design space with the
``best'' objective function value.  The fundamental issue here is how
to meaningfully compare two different mechanisms, which will generally
have incomparable performance.  For example, for two different
single-item auctions that depend on the valuation distributions
$\dist_1,\ldots,\dist_n$ only through $s$ samples
(Section~\ref{s:rev}), typically either one can have higher expected
revenue than the other, depending on the choice of $\dist_1,\ldots,\dist_n$.
Similarly, for two different combinatorial auctions with
low-dimensional bid spaces, one generally has higher welfare for some
valuation profiles, and the other for other valuation profiles.
The traditional approach in mechanism design to resolving such
trade-offs is to impose a prior on the unknown information and
maximize expected performance with respect to the prior.
But this approach would return us to the very bind we intended to
escape, of uninformatively complex optimal mechanisms that require
detailed distributional knowledge.

Are our insights surprising?
The presented results both confirm some existing intuitions --- which we
view as important sanity checks for the theory --- and go beyond them.
For example, in single-item auctions, the result that modest
data is sufficient for near-optimal revenue-maximization with
homogeneous bidders is natural given that the optimal auction depends
only on the valuation distribution's monopoly price.  While
revenue-maximization with heterogeneous bidders can only be a more
complex problem, it is not clear a priori how such complexity scales
with bidder heterogeneity, or even how ``complexity'' should be
defined.  The fact that the sample complexity scales linearly with the
number of distinct valuation distributions is a
satisfying and non-obvious formalization of the idea that ``heterogeneity
matters.'' 

For the case study of selling multiple items, the high-level
take-aways of our analysis are in line with prevailing intuition ---
simple auctions enjoy reasonable performance when there are no
complementarities among items, but not otherwise.
One pleasant surprise of the analysis, which deserves further
investigation, is that the positive results for simple auctions hold
even for 
the most general notion of  ``complement-free valuations,'' well
beyond the more well-studied special cases of gross substitutes and
submodular valuations.


\subsection{Open Questions}

This survey presented two recent applications of the approximately
optimal mechanism design paradigm.  There have been dozens of other
applications over the past fifteen years (Section~\ref{ss:refs}), and
there is still much to do.

For example, the sample-complexity formalism of Section~\ref{s:rev}
shows promise of deepening our understanding of Bayesian-optimal
mechanism design.
Proving that modest distributional knowledge suffices for near-optimal
mechanism performance is an important step in arguing the practical
relevance of a theoretically optimal design.
Upper bounds on the number of samples needed
(Section~\ref{ss:singlesample}) generally suggest interesting methods
of incorporating data, such as past bidding data, into designs.
Thus far, only the simple settings of single-item auctions
(Section~\ref{s:rev}) and single-bidder multi-item
mechanisms~\cite{DHN14} have been studied from this perspective.

For welfare-maximization with multiple items, results like those in
Section~\ref{s:wel} give preliminary insights into which auction
designs might work well, as a function of bidders' preferences.
An important direction for future work is to draw sharper distinctions
between different plausibly useful formats.
For example, there is ample empirical evidence that ascending auctions
for multiple items perform better than their sealed-bid counterparts.
Can this observation be made formal using the approximately optimal
mechanism design paradigm?


\begin{thebibliography}{10}

\bibitem{lovely}
Lawrence~M. Ausubel and Paul Milgrom.
\newblock The lovely but lonely {V}ickrey auction.
\newblock In Peter Cramton, Yoav Shoham, and Richard Steinberg, editors, {\em
  Combinatorial Auctions}, chapter~1, pages 57--95. MIT Press, Boston, MA, USA,
  2006.

\bibitem{BV03}
S.~Baliga and R.~Vohra.
\newblock Market research and market design.
\newblock {\em Advances in Theoretical Economics}, 3, 2003.

\bibitem{B99}
S.~Bikhchandani.
\newblock Auctions of heterogeneous objects.
\newblock {\em Games and Economic Behavior}, 26:193--220, 1999.

\bibitem{BK96}
J.~Bulow and P.~Klemperer.
\newblock Auctions versus negotiations.
\newblock {\em American Economic Review}, 86(1):180--194, 1996.

\bibitem{C+14}
I.~Caragiannis, C.~Kaklamanis, P.~Kanellopoulos, M.~Kyropoulou, B.~Lucier,
  R.~Paes Leme, and {\'{E}}.~Tardos.
\newblock On the efficiency of equilibria in generalized second price auctions.
\newblock arXiv:1201.6429, 2012.

\bibitem{CHK07}
S.~Chawla, J.~D. Hartline, and R.~D. Kleinberg.
\newblock Algorithmic pricing via virtual valuations.
\newblock In {\em EC}, pages 243--251, 2007.

\bibitem{C71}
E.~H. Clarke.
\newblock Multipart pricing of public goods.
\newblock {\em Public Choice}, 11(1):17--33, 1971.

\bibitem{CR14}
Richard Cole and Tim Roughgarden.
\newblock The sample complexity of revenue maximization.
\newblock In {\em STOC (to appear)}. ACM, 2014.

\bibitem{DRY10}
Peerapong Dhangwatnotai, Tim Roughgarden, and Qiqi Yan.
\newblock Revenue maximization with a single sample.
\newblock In {\em Proceedings of the 11th ACM Conference on Electronic Commerce
  {(EC)}}, pages 129--138, 2010.

\bibitem{DHN14}
S.~Dughmi, L.~Han, and N.~Nisan.
\newblock Sampling and representation complexity of revenue maximization.
\newblock Working paper, 2014.

\bibitem{feige}
Michal Feldman, Uriel Feige, Nicole Immorlica, Rani Izsak, Brendan Lucier, and
  Vasilis Syrgkanis.
\newblock A unified approach to valuations with restricted complements.
\newblock Working paper, 2014.

\bibitem{FFGL13}
Michal Feldman, Hu~Fu, Nick Gravin, and Brendan Lucier.
\newblock Simultaneous auctions are (almost) efficient.
\newblock In {\em 45th ACM Symposium on Theory of Computing (STOC)}, pages
  201--210, 2013.

\bibitem{G+06}
A.~V. Goldberg, J.~D. Hartline, A.~Karlin, M.~Saks, and A.~Wright.
\newblock Competitive auctions.
\newblock {\em Games and Economic Behavior}, 55(2):242--269, 2006.

\bibitem{G73}
T.~Groves.
\newblock Incentives in teams.
\newblock {\em Econometrica}, 41(4):617--631, 1973.

\bibitem{GS99}
F.~Gul and E.~Stacchetti.
\newblock Walrasian equilibrium with gross substitutes.
\newblock {\em Journal of Economic Theory}, 87(1):95--124, 1999.

\bibitem{hartline}
J.~D. Hartline.
\newblock Mechanism design and approximation.
\newblock Book draft. October, 2013.

\bibitem{HR09}
J.~D. Hartline and T.~Roughgarden.
\newblock Simple versus optimal mechanisms.
\newblock In {\em Proceedings of the 10th ACM Conference on Electronic Commerce
  {(EC)}}, pages 225--234, 2009.

\bibitem{H+11}
A.~Hassidim, H.~Kaplan, M.~Mansour, and N.~Nisan.
\newblock Non-price equilibria in markets of discrete goods.
\newblock In {\em Proceedings of the 12th ACM Conference on Electronic Commerce
  (EC)}, pages 295--296, 2011.

\bibitem{HMR14}
Z.~Huang, Y.~Mansour, and T.~Roughgarden.
\newblock Making the most of few samples.
\newblock Working paper, 2014.

\bibitem{KC82}
A.~S. {Kelso, Jr.} and V.~P. Crawford.
\newblock Job matching, coalition formation, and gross substitutes.
\newblock {\em Econometrica}, 50(6):1483--1504, 1982.

\bibitem{krishna}
V.~Krishna.
\newblock {\em Auction Theory}.
\newblock Academic Press, second edition, 2010.

\bibitem{LLN01}
B.~Lehmann, D.~Lehmann, and N.~Nisan.
\newblock Combinatorial auctions with decreasing marginal utilities.
\newblock {\em Games and Economic Behavior}, 55(2):270--296, 2006.

\bibitem{LOS02}
D.~Lehmann, L.~I. O'Callaghan, and Y.~Shoham.
\newblock Truth revelation in approximately efficient combinatorial auctions.
\newblock {\em Journal of the ACM}, 49(5):577--602, 2002.

\bibitem{LB10}
B.~Lucier and A.~Borodin.
\newblock Price of anarchy for greedy auctions.
\newblock In {\em Proceedings of the 21st Annual ACM-SIAM Symposium on Discrete
  Algorithms (SODA)}, pages 537--553, 2010.

\bibitem{milgrom}
P.. Milgrom.
\newblock {\em Putting Auction Theory to Work}.
\newblock Cambridge, 2004.

\bibitem{M81}
R.~Myerson.
\newblock Optimal auction design.
\newblock {\em Mathematics of Operations Research}, 6(1):58--73, 1981.

\bibitem{N03}
Z.~Neeman.
\newblock The effectiveness of {E}nglish auctions.
\newblock {\em Games and Economic Behavior}, 43(2):214--238, 2003.

\bibitem{amd_survey}
N.~Nisan.
\newblock Algorithmic mechanism design: Through the lens of multi-unit
  auctions.
\newblock In {\em Handbook of Game Theory}. North-Holland, 2014.
\newblock Forthcoming.

\bibitem{NR99}
N.~Nisan and A.~Ronen.
\newblock Algorithmic mechanism design.
\newblock {\em Games and Economic Behavior}, 35(1/2):166--196, 2001.

\bibitem{NS06}
N.~Nisan and I.~Segal.
\newblock The communication requirements of efficient allocations and
  supporting prices.
\newblock {\em Journal of Economic Theory}, 129(1):192--224, 2006.

\bibitem{OS09}
M.~Ostrovsky and M.~Schwarz.
\newblock Reserve prices in internet advertising auctions: A field experiment.
\newblock Working paper, December 2009.

\bibitem{R14}
T.~Roughgarden.
\newblock Barriers to near-optimal equilibria.
\newblock In {\em Proceedings of the 55th Annual IEEE Symposium on Foundations
  of Computer Science (FOCS)}, 2014.
\newblock To appear.

\bibitem{S03}
I.~Segal.
\newblock Optimal pricing mechanisms with unknown demand.
\newblock {\em American Economic Review}, 93(3):509--529, 2003.

\bibitem{comm_survey}
I.~Segal.
\newblock The communication requirements of combinatorial allocation problems.
\newblock In P.~Cramton, Y.~Shoham, and R.~Steinberg, editors, {\em
  Combinatorial Auctions}, chapter~11. MIT Press, 2006.

\bibitem{swinkels}
J.~M. Swinkels.
\newblock Efficiency of large private value auctions.
\newblock {\em Econometrica}, 69(1):37--68, 2001.

\bibitem{ST13}
V.~Syrgkanis and {\'E}.~Tardos.
\newblock Composable and efficient mechanisms.
\newblock In {\em Proceedings of the 45th ACM Symposium on Theory of Computing
  (STOC)}, pages 211--220, 2013.

\bibitem{V84}
Leslie~G. Valiant.
\newblock A theory of the learnable.
\newblock {\em Communications of the ACM}, 27(11):1134--1142, 1984.

\bibitem{V61}
W.~Vickrey.
\newblock Counterspeculation, auctions, and competitive sealed tenders.
\newblock {\em Journal of Finance}, 16(1):8--37, 1961.

\end{thebibliography}
\end{document}